\documentclass[oneside,11pt,reqno]{amsart}
 \pdfoutput=1
 \usepackage{geometry}                		
\usepackage{graphicx}				
\usepackage{amsmath}
\usepackage{amssymb}
\usepackage{hyperref}
\usepackage{color}
\usepackage{subcaption}
\usepackage{amsaddr}

\newcommand{\ket}[1]{\left| #1 \right>}

 \newcommand{\rmi}{\mathrm{i}}
 \newcommand{\rmd}{\mathrm{d}}
 \newcommand{\rme}{\mathrm{e}}
 \newcommand{\tr}{\mathrm{tr}}
 \newcommand{\Id}{\mathbb{I}}
\newtheorem{theorem}{Theorem}

\newcommand{\UN}{\mathrm{U}(N)}
\newcommand{\ONen}{\mathrm{O}^{-}(2N +2)}
\newcommand{\ONep}{\mathrm{O}^{+}(2N)}
\newcommand{\ONo}[1]{\mathrm{O}^{#1}(2N + 1)}
\newcommand{\SpN}{\mathrm{Sp}(2N)}

\newcommand{\abs}[1]{\lvert #1 \rvert}

\hypersetup{
    colorlinks=true,
    linkcolor=black,
    citecolor=black,
    filecolor=black,
    urlcolor=black,
}

\title[Fisher Hartwig determinants, CFT and universality]{Fisher Hartwig determinants, conformal field theory and universality in generalised XX models}

\author{J. Hutchinson}
\address{Department of Physics,\\ Indian Institute of Technology, \\Kanpur 208016, India}
\email{johutch@iitk.ac.in}
\author{N. G. Jones}
\address{School of Mathematics, \\University of Bristol, \\Bristol BS8 1TW, UK}
\email{n.g.jones@bristol.ac.uk}
\begin{document}
\begin{abstract}
We discuss certain quadratic models of spinless fermions on a 1D lattice, and their corresponding spin chains. These were studied by Keating and Mezzadri in the context of their relation to the Haar measures of the classical compact groups. We show how these models correspond to translation invariant models on an infinite or semi-infinite chain, which in the simplest case reduce to the familiar XX model. We give physical context to mathematical results for the entanglement entropy, and calculate the spin-spin correlation functions using the Fisher-Hartwig conjecture. These calculations rigorously demonstrate universality in classes of these models. We show that these are in agreement with field theoretic and renormalization group arguments that we provide. 
\end{abstract}
\maketitle

\section{Introduction}
Universality is a central concept in statistical physics, and in many contexts modern understanding is based on renormalization group (RG) arguments. Physical systems exhibit universal behaviour when they share an effective theory. This effective theory is reached by systematically removing irrelevant degrees of freedom, usually in the context of a flow of parameters in the space of Hamiltonians of quantum field theories. In this area rigorous arguments are scarce. In this paper, for a class of 1D spin systems which can be related to certain free fermion models, we present rigorously derived physical properties. We see that they correspond exactly to those expected from the analysis of the effective theory.  In particular, the systems we are interested in are critical -- the effective description is generically scale invariant and in fact has conformal symmetry, so can be described by a conformal field theory (CFT). We give the CFT description of our class of models, which takes the form of a direct sum of non-interacting massless fermions.  \par
Ground states of 1D critical models are characterised by algebraic decay of correlations and recently much interest has been shown in their entanglement structure. In \cite{CC,KC} it is seen that the von Neumann entanglement entropy of a subsystem of linear size $N$ diverges logarithmically as \begin{equation}\label{EE}S(N) = |A| \frac{c}{6} \log_2 N +{O}(1).\end{equation} Here $|A|$ is the size of the boundary of the subsystem, and $c$ is the central charge of the corresponding CFT. A similar formula for the XX model was observed numerically in \cite{VLRK} and then, including the extension to the R\'enyi entropies, was proved in \cite{JK} using the Fisher-Hartwig conjecture. This was then substantially generalised in \cite{KM,KM2} to wider classes of models susceptible to random matrix theory techniques. We review these results and see that \eqref{EE}, including the correction up to $o(1)$, as well as the expected spin correlations are verified exactly in these models.
\par
Our main results are the following:
\begin{itemize}
\item We give a physical interpretation of the large class of mathematical models studied in \cite{KM}.
\item This allows us to understand the universal entanglement entropy scaling found rigorously in that paper, and to predict the universal leading asymptotic scaling of equal time spin-spin correlation functions, as well as the oscillatory prefactors.
\item Where calculation is possible, we show that this prediction is indeed correct and find the coefficients of the leading terms using proven cases of the generalized Fisher-Hartwig conjecture. This corrects part of a similar calculation in \cite{HKM}.
\end{itemize}\par
The paper is organised as follows: in Section \ref{modeldefs} we introduce the lattice model that is the framework for our rigorous calculations. We then give a generic diagonal form of the Hamiltonian for the cases we will consider and analyse this diagonal form using standard quantum field theoretic methods. We aim to keep the discussion self contained so review some details.
We postulate the effective theory, and its dependence on microscopic parameters, and show that this is consistent using RG arguments. From this conformally invariant effective theory we can read off the central charge and give an argument, via bosonization, for the expected leading order behaviour of the correlation functions. \par

In Section \ref{groupmodels} we introduce in detail the three subclasses of the lattice model which permit further analysis and explain their physical relevance as well as the connection to the work of Keating and Mezzadri \cite{KM,KM2}. We then review the Fisher-Hartwig conjecture, proved in \cite{DIK} which is essential for the rigorous analysis that follows in Section \ref{RMT}.\par
We note that the bulk universality classes of our models, parameterised by a finite number of Fermi momenta and velocities, are identified in \cite{Korepin2} from the perspective of quantum integrable systems. The connection between integrals over classical compact groups and boundary conditions is also demonstrated in \cite{FF}. Finally, there is a complementary approach to identifying scaling dimensions and the central charge of the low energy theory via finite size scaling \cite{Cardy86,BIR,Korepin2} -- we will not pursue this, but note that it would also be accessible from the exact solution of our models.
\section{The quadratic fermion model -- definitions}\label{modeldefs}
We will consider the following model for free fermions on a lattice:
\begin{equation}
\label{ham}
H =  \sum_{j,k=0}^{M-1}
b^\dagger_jA_{jk}  b_k  -
2 h\sum_{j=0}^{M-1} b^{\dagger}_jb_j; \qquad\tr A=0
\end{equation}
which is equivalent, under Jordan-Wigner transformation, to the spin model 
\begin{equation}
\label{genmodel2}
\begin{split}
H & = -\frac{1}{2} \sum_{0 \le j \le k \le M-1}
A_{jk} (\sigma_j^x
\sigma_k^x+\sigma_j^y
\sigma_k^y) \left(\prod_{l=j+1}^{k-1}\sigma_l^z\right) -h\sum_{j=0}^{M-1}\sigma_j^z.
\end{split}
\end{equation}
The $b_i$ are the usual complex fermionic operators, and the $\sigma_i$ are the Pauli matrices. To fix conventions, the Jordan-Wigner transformation is given by:
\begin{align}
b^{\dagger}_{i} & = \frac{1}{2} (\sigma^{x}_{i} +i \sigma^{y}_{i})  \left(\prod^{i-1}_{l=1} \sigma^{z}_{l}\right);\nonumber \\
b_{i}&  = \frac{1}{2} (\sigma^{x}_{i} -i \sigma^{y}_{i})  \left(\prod^{i-1}_{l=1} \sigma^{z}_{l}\right);\nonumber \\
b_i^\dagger b_i &=\frac{1+\sigma_i^z}{2}. \label{JW}
\end{align}

\par The matrix $A$ is Hermitian and traceless, and without loss of generality we take it to contain only real parameters. We take periodic boundary conditions (PBCs) for the fermions\footnote{We note that this is inequivalent to periodic boundary conditions for the spins, but not drastically. Differences are of order $1/M$ as explained in \cite{LSM}, so in the thermodynamic limit we are calculating properties of the periodic spin chain.}, i.e. $b_M=b_0$. As with any quadratic fermion model, we can diagonalise by the methods of Lieb, Schulz and Mattis \cite{LSM} -- although we will see that after specialising to our cases of interest that a Fourier transform suffices. We will eventually take the thermodynamic limit $M\rightarrow \infty$. This system has a $\mathrm{U}(1)$ symmetry. For the spin model this corresponds to rotations in the XY plane: the action on the spin operators is conjugation by $R(\alpha) = \prod_{j}\exp(\rmi \frac{\alpha}{2}\sigma_z )$. For the fermions this is a uniform phase shift $b_j \rightarrow \exp(\rmi\alpha) b_j$.
In order to express the entanglement entropy and the correlation functions of the model in a way amenable to random matrix theory techniques we restrict the form of the matrices $A$ -- this is described in detail in Section \ref{groupmodels}. Essentially we restrict to either translation invariant (TI) chains, or to half chains where the TI bulk is broken by a boundary. In all cases we obtain, in the thermodynamic limit, a Hamiltonian of the form:
\begin{align}
H = \int_{-\pi}^{\pi} \rmd q \Lambda(q) \psi^\dagger(q) \psi(q) \label{Ham}
\end{align}
where $\psi(q)$ is a fermionic annihilation operator -- the Fourier transform of the lattice fermion annihilation operator $\psi_j$ -- and the energy is zero at the Fermi surface. As the system is non-interacting we have an exact single particle `Fermi sea' picture for our eigenstates, however $\Lambda(q)$ may be very general: we can construct any continuous periodic function that is symmetric in $q$. ({More precisely, an arbitrarily good approximation at the cost of longer distance interactions.}) \par
An important distinction should be made between two regions in parameter space -- the gapped region where $|\Lambda(q)|$ is strictly greater than zero and we have a ground state empty of particles, and a gapless region where $\Lambda(q)$ has zeroes. The gapless region is critical (it is at a quantum phase transition for an anisotropic variant of the model \eqref{ham} -- a generalised XY model -- defined in \cite{KM}) and is characterised by algebraic decay of correlation functions, as well as logarithmic scaling of entanglement entropy. We will focus our discussion on this gapless region.

\section{Physical picture}
We will continue schematically, working with the quadratic Hamiltonian \eqref{Ham}. The strategy is to isolate the low energy form of the Hamiltonian by linearisation and to see that this is consistent. More detailed accounts are given in, for example, \cite{Sachdev}; although much of the literature focuses on the nearest neighbour XX model: $A_{jk}=\delta_1(j-k)+\delta_{-1}(j-k)+U\delta_0(j-k)$, where $\Lambda(q) = \cos(q) + U$. The aim of this section is to see how the rigorous results for the entropy and the correlation functions given later can be argued from a physical perspective. In particular, for the groundstate of the system, we are interested in computing the leading asymptotic dependence of spin-spin correlation functions such as $\langle\sigma_x(N) \sigma_x(0)\rangle$ as $N\rightarrow \infty$, as well as identifying the central charge which controls the asymptotics of the entanglement entropy through \eqref{EE}.\subsection{Linearisation and irrelevance of local nonlinear corrections:}
\begin{figure}
\centering
\begin{minipage}{.5\textwidth}
  \centering
  \includegraphics[width=.9\linewidth]{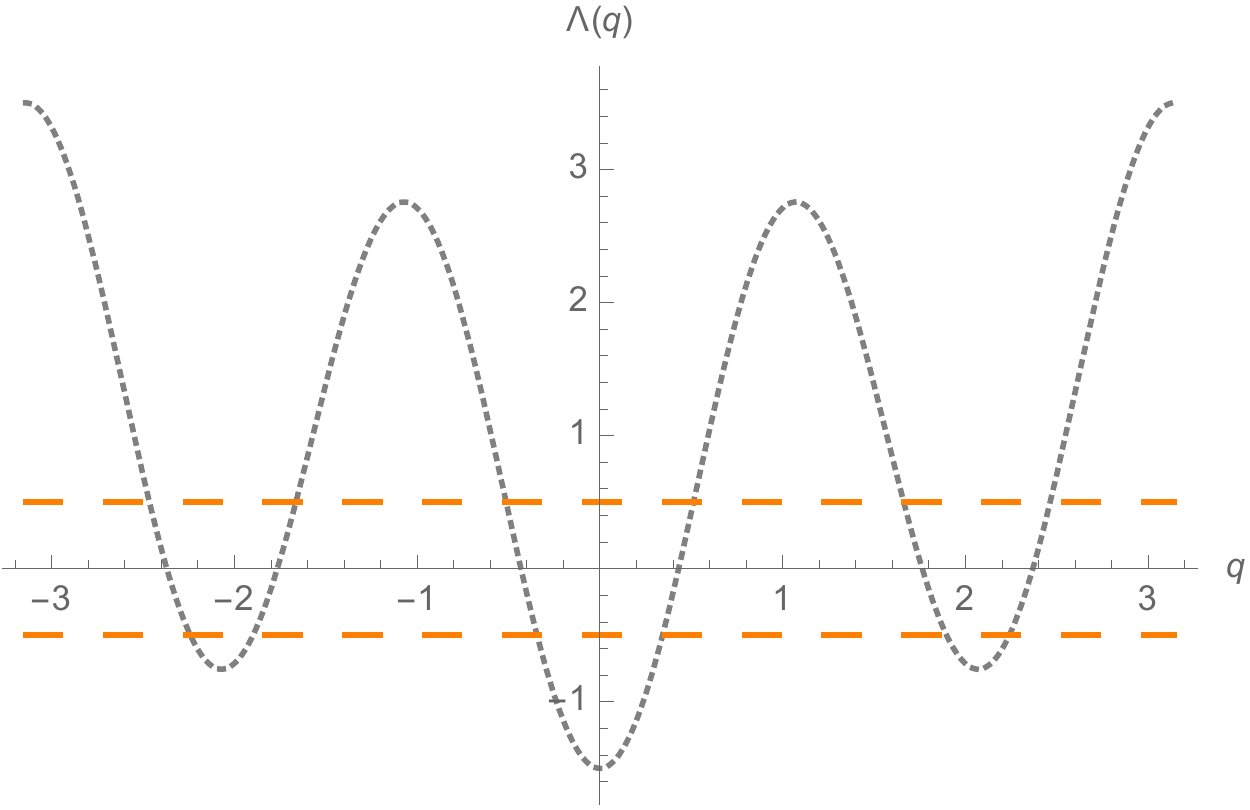}
  \captionof{figure}{Example of a dispersion with energy cutoff.}
  \label{disp}
\end{minipage}%
\begin{minipage}{.5\textwidth}
  \centering
  \includegraphics[width=.9\linewidth]{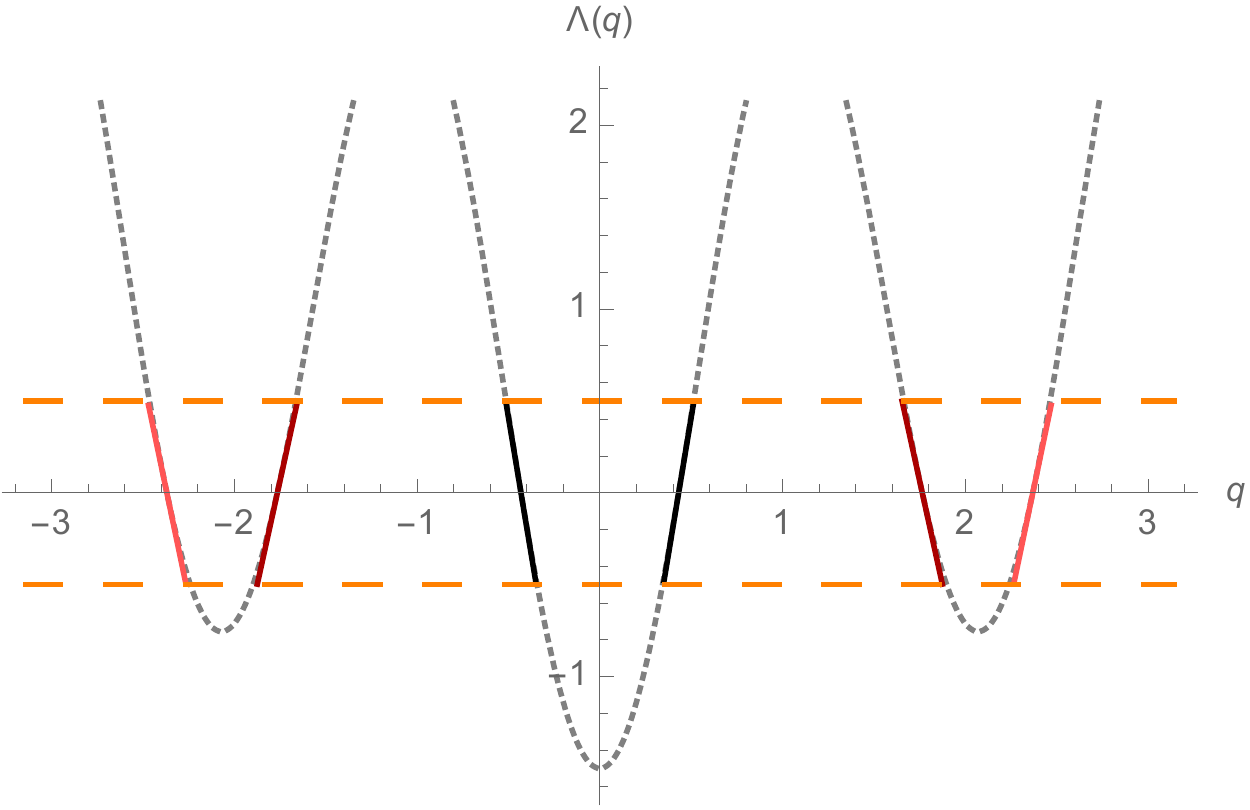}
  \captionof{figure}{The linearised dispersion.}
  \label{linear}
\end{minipage}
\end{figure}

We have an oscillatory dispersion which we may linearise about its zeroes (the 1D Fermi surface) to get:
\begin{align}H &\simeq \sum_{i=1}^{2R} v_i \int_{-\delta_i}^{\delta_i} \mathrm{d}q_i~ q_i ~\psi_i^\dagger(q_i)\psi_i(q_i)\nonumber\\
& = \sum_{i=1}^R |v_i| \int_{-\delta_i}^{\delta_i} \mathrm{d}q_i~ q_i ~(\psi_{i+}^\dagger(q_i)\psi_{i+}(q_i)-\psi_{i-}^\dagger(q_i)\psi_{i-}(q_i))\label{effham}\\
\label{fixedpoint} \mathrm{S}(\Delta) &= \sum_{i=1}^R  \int_{-\infty}^{\infty}\rmd \omega \int_{-\delta_i}^{\delta_i} \mathrm{d}q_i~ (-\rmi\omega+|v_i|q_i) \times\nonumber\\&
(\psi_{i+}^\dagger(\omega,q_i)\psi_{i+}(
\omega,q_i)-\psi_{i-}^\dagger(\omega,q_i)\psi_{i-}(\omega,q_i))
\end{align}
where $q_i$ is a coordinate around the $i$th point of the Fermi surface, which we will label $\theta_i$, and $ v_i = \frac{\partial }{\partial q} \Lambda(q)\rvert_{\theta_i}$ is the local Fermi velocity. $R$ is the number of zeroes between $(0,\pi)$. For the RG analysis, we consider the action $\mathrm{S}(\Delta)$, where $\omega$ is the Matsubara frequency, and where we put a uniform cutoff in the energy, $\Delta$, which locally restricts the momenta to an interval of width $2\delta_i = 2\Delta/|v_i|$. The action with unbounded $\omega$ would appear in path integral calculations of zero temperature correlation functions. As $\Lambda(q)$ is symmetric, we have pairs of points with equal and opposite velocity as illustrated in Figure \ref{linear}; we label these $\psi_{i+},\psi_{i-}$ for right and left movers respectively. The approximate (`low-energy') equality is because we are losing all details of the band curvature, and also, if we allow high energy excitation of the fields, we are overcounting degrees of freedom.
\par
\par
We note that it is possible that the linearisation fails -- this happens at points where both $\Lambda$ and a finite number of its derivatives vanish. We address this in detail in Appendix \ref{exc}, but for now we will limit our discussion to the generic case $\Lambda'(\theta_i)\neq0$.
\par
Corrections to the linearised Hamiltonian to account for the approximation do not affect the low energy physics in which we are interested. Following the discussion in \cite{Shankar,SCGP}, we first note that the action \eqref{fixedpoint} is a fixed point under the following RG transformation: we reduce the cutoff to ${\delta_i/s}$, for some $s>1$ to reach a new action $\mathrm{S}'(\Delta/s)$; then we rescale $(q,\omega,\psi_{i\pm}) \rightarrow (sq,s\omega,s^{-3/2}\psi_{i\pm})$ to restore the cutoff $\mathrm{S}'(\Delta)$. For the linearised action we indeed have $\rm S'= S$. The idea now is to see that our original action, before linearisation but with an energy cutoff, is in the basin of attraction of this fixed point. To do this we consider diagonal perturbations of the form: \begin{equation} \delta \mathrm{S} = \sum_{i,\pm}\int_{-\infty}^{\infty}\rmd \omega \int_{-\delta_i}^{\delta_i}  \rmd q_i \psi_{i\pm}^\dagger(\omega,q_i)\mu(\omega,q_i)\psi_{i\pm}(\omega,q_i).
\end{equation}
Expanding $\mu(\omega,q) = \sum_{j,k=0}^{\infty} \mu_{jk} (\rmi\omega)^jq^k$, we see that $\mu_{jk}$ scales as $s^{1-j-k}$. Hence as we reduce our cutoff, when $j+k\geq 1$ all such perturbations are irrelevant, the coefficient goes to zero. We are left with $\mu_{01}$ and $\mu_{10}$ which do not change under such a rescaling -- they are called marginal -- and their precise value is important. They correspond to the value of the Fermi velocity, which we fixed from our full Hamiltonian, and we see that this value does not change under the rescaling so does characterise different fixed points. Hence we have a different fixed point for each set of Fermi velocities, and we remark that this difference will manifest itself in the correlation functions. Finally the operator $\mu_{00}$ is transformed to $s\mu_{00}$, so grows as we reduce our cutoff -- it is relevant and can change which effective theory we need. This is a change in chemical potential, so naturally affects which effective theory we're looking at -- by moving the local\footnote{We are looking at each mode separately so can artificially rescale the chemical potential in each one.} zero of energy we can remove the degree of freedom entirely by going past a local minimum or maximum, or less drastically it will change the local Fermi velocity. As we have expanded around the true zeroes of our dispersion, we do not need need to consider perturbations of this type.
\par
The conclusion to take from this discussion is that the only important information from the full nonlinear Hamiltonian is the number of crossings, the Fermi momentum, $\theta_i$, at each crossing and the Fermi velocity at each crossing. All of the curvature is irrelevant as we look at low energies. We need the $\{\theta_i\}$ to ensure the correct charge density and relative phases between the different Fermi points. We then have universality classes of Hamiltonians which share these parameters. \par
We remark that we should work always with some cutoff so as not to allow overcounting of modes at high energy $q$ values, and we need not include couplings between the different modes as our full Hamiltonian is diagonal in $q$. We have neglected quadratic perturbations of the type $\psi^\dagger\psi^\dagger, \psi\psi$ which break the $\mathrm{U}(1)$ fermion number symmetry. Allowing such terms corresponds to considering anisotropic spin chains or superconducting fermionic chains -- these have a gap in the spectrum. Rigorous techniques are used in \cite{IMM,IJK} to calculate the leading order entanglement entropy for these systems, which is a constant. Finally we note that the analysis of interacting perturbations leads, as one would expect, to a much broader class of models with different physics to that described here -- however the rigorous techniques used below do not apply directly to such models so we will not discuss them further.  \par
\subsection{The central charge and entanglement entropy.}
Based on the above discussion, we fix the effective Hamiltonian as \eqref{effham}. To facilitate the CFT method, we note that:
\begin{align}
H_{CFT}= \sum_{i=1}^R |v_i| \int_{-\infty}^{\infty} \mathrm{d}q~ q ~(\psi_{i+}^\dagger(q)\psi_{i+}(q)-\psi_{i-}^\dagger(q)\psi_{i-}(q))\label{hamC}\end{align}
is clearly in the same low energy universality class (we have just extended the high energy cutoff) and should have the same long distance physical properties. The advantage is that now we have a real space continuum conformal field theory, corresponding to the continuum limit\footnote{Roughly, after defining a lattice spacing $a$, let $\psi(x) = \lim_{a\rightarrow0}\frac{1}{\sqrt{a}}b_{x/a}$. In our microscopic models factors of $a$ are suppressed throughout. Care should be taken in the presence of a boundary, as the lattice may need to be shifted before this identification.} of our chain, and can use textbook methods. We maintain a fixed density of fermions in our ground state.

We should pay attention to the boundary conditions of the chain that led to \eqref{Ham}, and carry them over to \eqref{hamC}. We note that we have a direct sum of $2R$ chiral fermions, each with a central charge of $1/2$ \cite{diF}. In total the model has $c = R$. Hence on the basis of \cite{CC} we expect the entanglement entropy for a system with PBCs to have the entanglement entropy of any block of length $N$ to scale as:
\begin{equation}
S(N) = \frac{R}{3}\log_2 N + \kappa_\Lambda + o(1).\label{UNpre}
\end{equation}
For a system with a boundary, irrespective of the boundary condition, we expect:
\begin{equation}
S(N) = \frac{R}{6}\log_2 2N + \frac{1}{2}\kappa_\Lambda + (\log_2(g)) +o(1),\label{ONpre}
\end{equation}
where we consider blocks of length $N$ starting at the boundary. (This choice of block is motivated by the random matrix theory calculations below - we note that \cite{FC} discusses blocks away from the boundary in that language.)\par
To interpret these formulae, first we see that the constant $\kappa_\Lambda$ is not universal, but agrees for systems with different boundary conditions sharing a (proportional) bulk dispersion relation $\Lambda(q)$. Going between \eqref{UNpre} and \eqref{ONpre} there is a factor of two separating the `bulk terms' originating in the fact that our block from site 1 to site $N$ has gone from having two bulk-facing edges to one (this is the $|\mathcal{A}|$ in \eqref{EE}). The scaling by $2N$ inside the logarithm is a result of a method of images argument given in \cite{CC} - broadly: studying $N$ sites with a boundary can be mapped into $2N+1$ sites without a boundary, we will see this in more detail below. We include the bracketed $\log_2(g)$ for completeness. This is the boundary entropy introduced in \cite{AL}, where it is also shown that for free fermions $g=1$ -- hence this term will be zero for our systems.\par
\subsection{Spin correlation functions}  
We can use the effective theory \eqref{hamC} to find the asymptotics of the spin-spin correlation functions by bosonization methods \cite{Sachdev,DS2,DS,McG,FG,Ov}. The aim of this section is to motivate an expected leading asymptotic term, rather than make an explicit first principles derivation.\par
 We defer to the literature for detailed definitions and discussion, for our purposes we need the bosonization formula\footnote{All expressions are subject to appropriate normal ordering. We suppress the Klein factors which fix the correct mutual anticommutation relations between the fermionic operators as they do not play a role in our calculations.} \cite{FG}:
\begin{align}\label{bosonize}
\psi_{i}(x) \simeq \exp\rmi \Phi_{i}(x)\biggl(A_{i+}\exp\rmi(\vartheta_{i}(x)+ \theta_i x)+A_{i-}\exp-\rmi(\vartheta_{i}(x)+ \theta_i x)\biggr); 
\end{align}
where $\Phi_i$, $\vartheta_i$ are canonical free, massless, bosonic fields which bosonize each of the $R$ complex fermions $\psi_i$. For each fermion $\theta_i$ is the Fermi momentum and the $A_i$ are constants. Again, we have a `low energy' equality, i.e. this is the dominant part of the operator when calculating long distance correlation functions.
Physically $\exp(\rmi \Phi_i$) creates a unit charge excitation, while $\pm(\theta_i+\partial_x\vartheta_i)/\pi$ is the relevant charge density $\partial Q_i/\partial x$ -- the expression above is thus analogous to the Jordan-Wigner transformation \eqref{JW} for these continuum fields.  For convenience we will also define the left and right moving boson fields:
\begin{align}
\varphi_{i\pm}=\Phi_i \pm \vartheta_i.
\end{align}\par
Now we argue on general grounds that, to leading order as $x\rightarrow\infty$, the spin operator must take the form:
\begin{align}
\sigma^+(x)= (-1)^{bx} \sum_{\tilde\beta} C_{\tilde\beta} \exp\rmi\left(\sum_{j=1}^R \tilde\beta_j (\varphi_{j+}(x)+Q_j(x)) + \tilde\beta_{2R+1-j} (\varphi_{j-}(x)-Q_j(x)) \right) \label{vertexspinop}
\end{align}
where we consider all strings $\tilde\beta_j = \pm 1/2$ such that $\sum_j \tilde\beta_j = 1$. $Q_j(x)$ is the cumulative charge to the left of $x$ for the $j$th fermion and $C_{\tilde\beta}$ are constants. The N\'eel order factor $(-1)^{bx}$ for $b$ either 0 or 1 is included for consistency with the lattice spin operator and the Jordan-Wigner transformation\footnote{In any particular lattice model this term is removable by a redefinition of the on-site spin basis. Starting from a fixed basis it should in principle be derived from treating boundary conditions and branch cuts in the field theory carefully, however we will not pursue this here.}.
\par
First, we expect by analogy with the single complex fermion XX model case \cite{Sachdev} that we will deal with vertex operators -- exponentials of our left and right moving bosonic fields. Vertex operators give local expressions for certain non-local operators in the fermionic degrees of freedom, and the continuum limit of the spin operator must be among these. Such operators take the form $V_{\alpha,\gamma}: = \exp\rmi \sum_i (\alpha_i \varphi_{i+} + \gamma_i \varphi_{i-})$. In order for $V_{\alpha,\gamma}$ to be well defined, we must have that $\alpha_i -\gamma_i \in \mathbb{Z}$ and $\alpha_i +\gamma_i \in \mathbb{Z}$.
The correlation functions of such operators are easily derived, giving:
\begin{align}
\langle V_{\alpha,\gamma}(x)V_{\alpha',\gamma'}(0)\rangle = \frac{\delta_{\alpha+\alpha',0}\delta_{\gamma+\gamma',0}}{x^{\frac{1}{2}\sum(\alpha_i^2+\gamma_i^2)}}. \label{vertexcorr}
\end{align}
Clearly, then, the dominant vertex operators in the large $x$ limit have either $\alpha_i = \gamma_i = 0$ or $\alpha_i,\gamma_i = \pm 1/2$. The lattice spin operator satisfies nontrivial commutation relations which must carry over to the continuum in each fermion sector, hence we can exclude the first case. This gives us that, to leading order, all $\tilde\beta_j = \pm1/2$ in the spin operator above. The condition $\sum_j \tilde\beta_j =1$ is due to the constraint that the spin operator creates a single particle, or equivalently that it transforms under the $\mathrm{U} (1)$ rotation symmetry in the same way as a single fermion. The condition enforces this by noting that for $\psi_i \rightarrow \exp(\rmi\alpha)\psi_i$, we have $\Phi_i \rightarrow \Phi_i +\alpha$ and $\vartheta_i \rightarrow \vartheta_i$. Finally, every appearance of $\vartheta_i$ is through integrating the charge density and so should be accompanied by $Q_i(x)$ -- the form of \eqref{vertexspinop} ensures this. The bosonization method cannot give the coefficients so we assume that they are nonzero.\par
Then, using \eqref{vertexspinop} and \eqref{vertexcorr}, we can give the expected form of the asymptotics for our lattice spin-spin correlators as:
\begin{align}
\langle\sigma^+(N)\sigma^-(0)\rangle &\sim (-1)^{bN} N^{-R/2}\sum_{\tilde\beta} |C_{\tilde\beta}|^2 \exp\rmi\left(\sum_{j=1}^R (\tilde\beta_j - \tilde\beta_{2R+1-j})Q_j(N) \right)\\
&\sim (-1)^{bN}N^{-R/2}\sum_{\tilde\beta} |C_{\tilde\beta}|^2 \exp\rmi N\left(\sum_{j=1}^R (\tilde\beta_j - \tilde\beta_{2R+1-j})\theta_j \right).\label{fieldcorr}
\end{align}
This gives us $\Re\langle\sigma^+(N)\sigma^-(0)\rangle/2=\langle\sigma_x(N)\sigma_x(0)\rangle=\langle\sigma_y(N)\sigma_y(0)\rangle$ by the isotropy of the Hamiltonian (and that there can be no spontaneous breaking of this continuous rotation symmetry in the ground state by the result of \cite{coleman1973}). We note that for every allowed string $\{\beta_j\}$ in the sum there is a corresponding string $\{\beta_j'\}$ with the opposite angle in the exponential -- if we assume $|C_\beta| = |C_{\beta'}|$ then the right hand side of \eqref{fieldcorr} is real. This property is verified in the exact calculation below. Finally we note that physically this operator is a sum of operators, each of which create and annihilate charges (always with an overall +1) and create particle-hole excitations. When $R$ is odd we have some non-oscillating terms which contain only the charged $\Phi_i$ fields.\par
For completeness, we have from \eqref{JW} that $\sigma_z$ on the lattice is, up to a constant shift, the fermion number operator. Hence the analogous continuum operator is $\sigma_z(x)\sim \sum_i\partial_x \vartheta_i$ from which we quickly see that $\langle\sigma_z(x)\sigma_z(0)\rangle \sim x^{-2}$.\par
 We implicitly assumed above that we were working with periodic spatial boundary conditions. We expect the same asymptotic correlations in the presence of a boundary, via a method of images argument as in \cite{diF}. Note that in the presence of a boundary we must have a conformally invariant boundary condition, for example $\psi(x)=0$ or $\psi(x)=\infty$, which respectively correspond to the `ordinary' and `extraordinary' transitions discussed later.\par 

\section{Models related to classical compact groups}\label{groupmodels}
We will now specialise $\eqref{ham}$ to three models which encompass all the models labelled by the classical compact groups introduced in \cite{KM}.
In the Hamiltonian \eqref{ham} we separate for clarity of interpretation the kinetic term with hopping matrix $A$ from the term proportional to $h$ which acts as a chemical potential for the fermions, or as an external field for the spins. Below, though, we will summarise the whole Hamiltonian by  $\overline{A}=A-2h\Id $.
\subsection{Translation invariant chain}\label{TI}
Let: $\overline{A}_{jk} = a(j-k)$ where $a$ is an even function on $\mathbb{Z}/M\mathbb{Z}$. Let $M$ be odd, even $M$ is very similar with formulae given in \cite{KM}. Defining 
\begin{equation}
\delta_i(j) = \left\{
        \begin{array}{ll}
            1 & \quad i = j \mod M\\
            0 & \quad i \neq j \mod M
        \end{array}
    \right.\end{equation}
    we have: \begin{equation} a(j) = \alpha_0\delta_0(j) + \sum_{i=1}^{(M-1)/2}\alpha_i(\delta_i(j)+\delta_{-i}(j)). \end{equation}
    We can diagonalise the Hamiltonian by Fourier transform to get:
    \begin{align}
    H &= \sum_q \Lambda_q b^\dagger_q b_q; \qquad \mathrm{for} \qquad
    \Lambda_q= \alpha_0 + \sum_{j=1}^{(M-1)/2} 2\alpha_j\cos(q j).
    \end{align}
   Now taking the limit $M\rightarrow \infty$ we get $\Lambda(q) = \alpha_0 + 2\sum_{j=1}^{\infty} \alpha_j \cos(jq)$ where $\alpha_j$ decays sufficiently fast that this series converges. This decay is physically natural as $\alpha_j$ is a coupling of sites $j$ lattice spacings apart and so implies that our Hamiltonian is local in both the fermion and spin operators. In this limit we have also, after redefining particle/hole excitations relative to the Fermi level:
   \begin{equation}
\psi^\dagger(q) := 
\begin{cases}
b^\dagger_q, & \Lambda(q)>0 \\
b_q, &\Lambda(q)<0,\end{cases}
\end{equation}
the Hamiltonian:
\begin{align}
H = \int_{-\pi}^{\pi} \rmd q |\Lambda(q)| \psi^\dagger(q) \psi(q).
\end{align}
This is the $\UN$ model of \cite{KM} -- the translation symmetry of the chain corresponds to the translation symmetry of the Haar measure on $\UN$. We will see later that the entanglement entropy and correlation functions are given by Toeplitz determinants which can be written as integrals over $\UN$.

\subsection{Semi-infinite chain -- ordinary transition}\label{ord}
We will now work directly in the thermodynamic limit, however specialisation to the finite chain is similar and described in, for example, \cite{FC} where random matrix techniques are used to study the R\'enyi entanglement entropies in the XX model on open chains. The following is close to the method of images from differential equation theory and also to the same method in CFT with a boundary as explained in \cite[Chapter 11]{diF}.\par
Consider the Hamiltonian for fermionic operators $\{B_j\}_{j\in\mathbb{Z}_+}$ on the semi-infinite chain:
\begin{equation}
H'= \sum_{j=1}^{\infty} B^\dagger_{j+1} B_{j} + h.c.~, \label{halfham}
\end{equation}
where h.c. stands for the hermitian conjugate. This is the nearest neighbour fermionic chain with open boundary conditions (OBCs). We will use this model as an example to illustrate the main ideas of the calculations given in detail in Appendix \ref{images}.
To diagonalise this model we introduce the `mirrored' operators $B_{-j}:=-B_j$. Then\footnote{We also include $B_0 = - B_0 = 0$. The factor of two in front of the dispersion is due to the density of modes and is explained in Appendix \ref{images}.}:
\begin{equation}
H'= \frac{1}{2}\sum_{j=-\infty}^{\infty} B^\dagger_{j+1} B_{j} +h.c. = 2\int_{-\pi}^{\pi}\rmd q\cos(q) B^\dagger(q)B({q})  ~.
\end{equation}
where the Fourier transformed operators $B(q) \propto \sum_{j=-\infty}^{\infty} \exp(\rmi qj)B_j$ or, in terms of the original, unmirrored variables, $B(q)\propto \sum_{j=1}^{\infty} \sin(q j) B_j $.\par
To make contact with the models and methods of \cite{KM,HKM} and the canonical Hamiltonian \eqref{ham}, suppose we start with fermions $\{b_j\}_{j\in\mathbb{Z}}$ on the whole chain. We can use these to simulate the above half chain by introducing the parity odd combination: $B_j := (b_{j} - b_{-j})/\sqrt 2$. Inserting this into the Hamiltonian \eqref{halfham} we get:
\begin{align}
H'&= \frac{1}{2}\sum_{j=1}^{\infty} (b_{j+1}^\dagger-b_{-(j+1)}^\dagger)(b_j-b_{-j})  + h.c. \\
&= \frac{1}{2}\sum_{j=-\infty}^{\infty} b_{j+1}^\dagger b_j -\frac{1}{2}\sum_{j=-\infty}^{\infty} b_{j+1}^\dagger b_{-j} + h.c.~. \nonumber
\end{align}
In the $b_j$ degrees of freedom we now have a non-local Hamiltonian, with hopping between sites $i$ and $j$ if $|i-j|=1$ or if $|i+j| = 1$. As a final step we shift the chain by one lattice site to align with the model \eqref{ham} which starts at $j=0$ (with a $b_0$ not identically 0) and so send $j\rightarrow j-1$. Hence we see \eqref{halfham} is a particular case of \eqref{ham} with $A_{jk} = \delta_1(|j-k|) - \delta_1(|j+k+2|)$. This is in turn a particular case of the model labelled $\SpN$ in \cite{KM}, with hopping matrix given in Table \ref{models}. Note that combining the steps above we also have the spectrum and eigenfunctions of this particular model.\par 
We now assert, with details left to  Appendix \ref{images}, that all of the nonlocal models in \cite{KM} that are related to the groups $\SpN,\ONen,\ONo{+}$ are more naturally viewed as local models on the half chain, after mirroring with the parity transformation $B_j = -B_{-j}$. These models are defined by the hopping strengths $a(j-k)$ in: \begin{equation}
H'= \sum_{j,k=1}^{\infty} a(j-k)B^\dagger_{k} B_{j} + h.c. + H_{edge}~,
\end{equation}
where $H_{edge} =  - \sum_{j,k=1}^{\infty} a(j+k)B^\dagger_jB_k$. $H_{edge}$ is localised near the boundary (due to the decay of $\alpha_j$) and is needed so that the transformation to the full chain yields a translation invariant Hamiltonian. Note that this gives an overall weakening of hopping amplitude near the boundary. Such models have $\langle B_0^\dagger B_0\rangle = 0$ by definition, and correspond to the `ordinary' transition for our critical system, as described in \cite{Cardy,diF}. Ideas from boundary CFT then suggest that the long distance correlation functions will be identical to those of the infinite chain with a hopping matrix $A_{jk}=a(j-k)$. 
\subsection{Semi-infinite chain -- extraordinary transition} It is then natural to consider the alternative parity transformation $B_{-j}:=B_j$  and $H_{edge} =  {+}\sum_{j,k=1}^{\infty} a(j+k)B^\dagger_jB_k$. The analysis goes through mostly as before, with details in Appendix \ref{images}. The dispersion relation is identical to that of the models in \ref{ord}.\par Importantly we no longer have $B_0 = 0$. In fact for finite chain size $\langle B_0^\dagger B_0\rangle \sim N_e/M$, the filling fraction (which will be finite in the thermodynamic limit). We have enhanced hopping amplitudes near the boundary, giving a finite order parameter there at the transition (again see \cite{Cardy,diF}). Thus we are now in models representing the `extraordinary' transition of a system with a bulk corresponding to some $\UN$ model. As a final remark we note that the nonzero expectation value here will manifest itself as the conformally invariant $\psi(0) = \infty$ boundary condition in the continuum limit, due to scaling by inverse powers of the lattice constant in the definition of $\psi(x)$.
\subsection{Summary}
\begin{table}
\centering
\begin{tabular}{c c c c}
\hline \hline
Physical & Classical compact  & Structure of matrices & Matrix entries \\
model & group & $\overline{A}_{j,k} $ & $\left( \mathcal{M}_{N} \right)_{j,k} $ \\
\hline \hline
TI, PBC &$\UN$ & $a(j-k) $ & $f_{j-k}, \quad j,k \geq 0$ \\
TI, Edge, $P = 1$&$\ONep $ & $a(j-k)+a(j+k)$ & $f_{0} \quad$ if $j=k=0$\\
& & & $\sqrt{2} f_{l}$  if \\
&& &  either $j=0, k=l$ \\
&& & $\quad$ or $j=l, k=0$ \\
 && & $f_{j-k} + f_{j+k}, \quad j,k >0$ \\
TI, Edge, $P=-1$&$\SpN$,& $a(j-k) -a(j+k+2)$ & $f_{j-k}-f_{j+k+2}, \quad j, k \geq 0$ \\
&\quad$\ONen$ & & \\
TI, Edge, $P=\mp 1$&$\mathrm{O}^{\pm}(2N+1)$ & $a(j-k) \mp a(j+k+1)$ & $f_{j-k} \mp f_{j+k+1}, \quad j,k \geq 0$\\ 
\hline
\end{tabular}
\caption{Entries of matrices $\mathbf{\bar{A}} = \mathbf{A}- 2 h \mathbf{I}$ in terms of the function $a(j)$. The $f_{k}$s are the Fourier coefficients of the symbol $f\left( \theta \right)$ of $\mathcal{M}_{N}$ (see Section \ref{RMT}). $P$ corresponds to the parity of the operators $B_j \propto b_j \pm b_{-j}$, which in turn indicate the nature of the boundary transition.}
\label{models}
\end{table}
In Table \ref{models} we summarise all of these models. First we identify the physical model with a TI bulk, which can be easily transferred to a predictive CFT description. Then we give the corresponding compact group model of \cite{KM}, which is fully specified by the $\overline A$ matrix in the Hamiltonian \eqref{ham} and allows rigorous analysis via random matrix theory methods. We denote the different connected components of the orthogonal group by $\ONep$, $\ONo{+}$, $\ONo{-}$ and $\ONen$ with plus and minus corresponding to positive and negative determinant respectively.
\section{Asymptotics for determinants of Toeplitz matrices}
\label{toe}
We use this section to recall some results on the generalised Fisher Hartwig conjecture for asymptotics of certain Toeplitz matrices, and related Toeplitz+Hankel matrices. We note first that a Toeplitz matrix has the form $M_{jk}=M_{j-k}$ and a Hankel matrix has the form $M_{jk}=M_{j+k}$.
 We will make use of these results in the following section to compute both the entanglement entropy and the correlators. Originally conjectured by Basor and Tracy \cite{BT}, Dieft, Its and Kravsovsky \cite{DIK} proved the Fisher Hartwig conjecture for the situations we will now describe. We note that these determinants are often equivalent to integrals over random matrix distributions, as discussed in \cite{KM,FF}. \par
We wish to compute the determinants of matrices with Toeplitz structure, with entries given by the Fourier coefficients of a function $f(z)$ called the symbol (or generating function);
\begin{equation}
D_{N}(f(z)) = \det (f_{j-k})_{j,k=0}^{N-1}, \qquad f_{j} = \frac{1}{2\pi} \int^{2 \pi}_{0} f(\theta) e^{- i j \theta} d \theta.
\end{equation}

We will be interested only in symbols whose $m$ singularities are jump discontinuities, as in Figure \ref{symbolfig}, thus $f(z)$ can be written in following form on the unit circle:
\begin{equation}
f(z) = e^{V(z)} z^{\sum_{j=1}^m \beta_j}\prod_{j=1}^m g_{z_j,\beta_j}(z) z_j^{-\beta_j} \qquad z=e^{i\theta}, \qquad \theta \in [0, 2\pi),
\label{fz}
\end{equation}
where $z_{j} = e^{i \theta_{j}}$, $\beta_{j} \in \mathbb{C}$ for $j=1, \ldots m$, $0 = \theta_{1} < \ldots < \theta_{m} < 2 \pi$,
\begin{equation}
g_{z_{r},\beta_{r}} (z) = 
\begin{cases}
e^{i \pi \beta_{r}}, & 0 \leq \mathrm{arg} z < \theta_{r}, \\
e^{-i \pi \beta_{r}}, & \theta_{r} \leq \mathrm{arg} z < 2 \pi, \\
\end{cases}
\end{equation}
and the function $V(e^{i \theta})$ satisfies some smoothness conditions, explained in \cite{DIK}.
\par
In general there can be multiple ways of writing this decomposition, each giving different representations of $f(z)$. We will use the nomenclature of \cite{DIK} and define these other representations relative to an initial one in the following way. Start with one such representation, in terms of a set $\{\beta_{j}\}$, and then modify the right hand side of \eqref{fz} by $\tilde\beta_j = \beta_j +n_j$, subject to the constraint
\begin{equation}
\sum^{m}_{j=1} n_{j} = 0.
\label{nj}
\end{equation}
This is called a \emph{Fisher-Hartwig representation} $f(z;n_{1}, \ldots n_{m})$, relative to the initial $f(z;0,\ldots,0)$. Importantly we note that our initial symbol is given by: \begin{equation}f(z) = \prod_{j=1}^mz_j^{n_j} f(z;n_{1}, \ldots n_{m}). \end{equation}\par

To find the leading term of the asymptotics of the determinant we need the subset of these representations which minimise $\sum_{j=1}^{m} \tilde\beta^{2}_{j}$. For symbols that are even and with image $\{+\exp(\rmi\phi),-\exp( \rmi\phi)\}$, for some $\phi \in\mathbb{R}$, (featuring prominently below) we find that these minimal representations have $\tilde\beta_{j} = \pm \frac{1}{2}$ and $n_{j}\in \{1,-1,0\}$. \par
We now quote the following theorem, after simplifications which occur in our cases of interest. In particular, our symbols will always have $V(z)=$ constant, so the only Fourier coefficient is $V_0$. 
\begin{theorem}[Deift, Its, Krasovsky 2011]\label{FHT}
As the matrix dimension, $N$, goes to infinity:
\begin{align}
\det(f_{j-k})_{j,k=0}^{N-1}=\sum_{\mathrm{Reps:}~\{n_j\}}\left(\prod_{j=1}^m z_j^{n_j}\right)^N
\mathcal{R}(f(z;\{n_j\})(1+o(1)).\end{align}
Where:\begin{align}
\mathcal{R}(f(z;\{n_j\}))&=N^{-\sum_{j=1}^m \tilde\beta_j^2}\exp\left(NV_0\right) \nonumber
\prod_{1\le j<k\le m}
|z_j-z_k|^{2\tilde\beta_j\tilde\beta_k}
\prod_{j=1}^mG(1+\tilde\beta_j) G(1-\tilde\beta_j)\\
\tilde\beta_j&=\beta_j+n_j .\nonumber
\end{align} \end{theorem}
We emphasise that $V_0$ is unaltered when passing between representations. $G(z)$ is the Barnes G-function \cite[\textsection 5.17]{NIST:DLMF}. For imaginary $V_0$, the dominant behaviour of this determinant is an algebraic decay. We also note that, using the methods of \cite{DIK}, one can write Hankel determinants and Toeplitz+Hankel determinants in terms of purely Toeplitz determinants which can then be put into the full version of the above theorem -- we return to this point below.

\section{Random matrix theory results}\label{RMT}
\begin{figure}
\centering
\begin{minipage}{.5\textwidth}
  \centering
  \includegraphics[width=.9\linewidth]{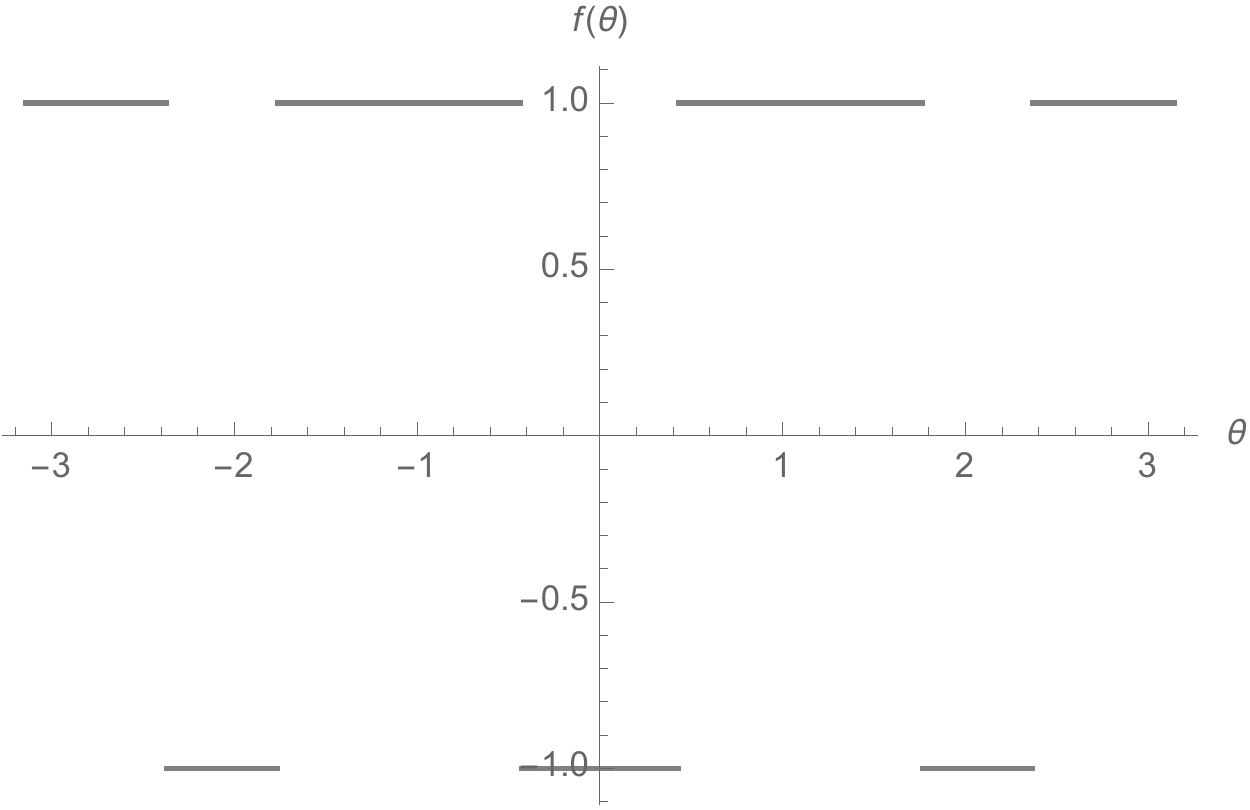}
  \captionof{figure}{The symbol for the dispersion in Figure \ref{disp}}
  \label{symbolfig}
\end{minipage}%
\begin{minipage}{.5\textwidth}
  \emph{Example:}\\ 
  In Figure \ref{disp} the dispersion is given by:
  $$\Lambda(\theta)=1 - \cos(\theta)/2 - 2 \cos(3\theta),$$
  and the symbol in Figure \ref{symbolfig} is: $$f(\theta)=\mathrm{sign}(\Lambda(\theta)).$$
  The symbol has six jump discontinuities, so we have $R=3$. The only other data required for the Fisher-Hartwig analysis are the values of $\theta$ at the jumps and that $f$ is negative at zero. We use periodicity to shift the domain to $[0,2\pi)$.
   \end{minipage}
\end{figure}\par

We see from the results of \cite{KM,HKM} that for all models discussed in this paper, the asymptotics of both the entanglement entropy and certain spin-spin correlation functions may be related to determinants of combinations of Toeplitz and Hankel matrices. The relevant matrix structure is dictated by Table \ref{models} and the entries of these matrices are given in terms of Fourier coefficients of the symbol:
\begin{equation}
f(\theta) = \frac{\Lambda(\theta)}{|\Lambda(\theta)|}= \mathrm{sign} (\Lambda (\theta))\label{symb}.
\end{equation}
This is an even function with jump discontinuities taking values $\pm 1$. An example is given in Figure \ref{symbolfig}. $\Lambda (\theta)$ is the $M \rightarrow \infty$ limit of the dispersion relation and $d$ is a constant, which we see from Table \ref{models} will take values either $0$, $1$ or $2$ for the systems we are concerned with. We remind the reader that $R$ is the number of zeroes of $\Lambda$ in $(0,\pi)$ and $\theta_i$ are the points where $\Lambda(\theta_i) = 0$ (and the first nonzero derivative of $\Lambda$ is odd). The number of jump discontinuities of our symbol is given by $m=2R$. We note that the fermionic correlators, and hence the $\sigma_z$ two point functions, are simply related to the Fourier coefficients themselves. 
\subsection{Entanglement}
We begin by stating the results from the entanglement computation in \cite{KM}. Thanks to the connection between the reduced density matrix and ground state correlation functions for one dimensional free fermions noted in \cite{Peschel,VLRK}, we can relate the entanglement entropy of such a system to the `correlation matrix' - the eigenvalues of which are simply related to the fermionic correlations \emph{within} the subsystem.
More precisely, the entanglement entropy of a subchain of length $N$ can be shown to be given by the contour integral:
\begin{equation}
\label{JKidea}
S(N) = \lim_{\epsilon \rightarrow 0^+}
\lim_{\delta \rightarrow 0^+} \frac{1}{2\pi \rmi}
\oint_{c(\epsilon,\delta)} \rme(1 + \epsilon,\lambda) \frac{\rmd
\ln D_N[f](\lambda)}{\rmd \lambda} \rmd \lambda.
\end{equation}
This is a result of the residue theorem, as this integral sums the bipartite entanglement function:\begin{equation}
\label{binaryent}
\rme(x,\nu)= - \frac{x +
\nu}{2}\log_2\left(\frac{x + \nu}{2}\right) - \frac{x -
\nu}{2}\log_2\left(\frac{x - \nu}{2}\right)\end{equation}
at each of the eigenvalues of the correlation matrix - which are themselves zeroes of the characteristic polynomial:
\begin{equation}
D_N[f](\lambda) = \det\left(\lambda \mathbb{I} - \mathcal{M}_{N}[f]\right).
\end{equation}
Physics constrains the eigenvalues to lie in $[-1,1]$, hence we take $c(\epsilon,\delta)$ to be any curve enclosing the interval $[-1,1]$ and approaching it as $\epsilon,\delta\rightarrow 0$. $\mathcal{M}_N$ is the relevant Toeplitz+Hankel matrix of Table \ref{models}. Note that for the TI chain: $\mathcal{M}_{N}[f]_{jk} = f_{j-k}$ with symbol defined by \eqref{symb} -- thus $D_N[f](\lambda)$ is a Toeplitz determinant and has leading asymptotics given by the relevant extended form of Theorem \ref{FHT} given in \cite{KM}. For the other groups a different proven case of the Fisher-Hartwig conjecture was used \cite{FF}, in order to deal with Toeplitz+Hankel structures. \par By analysing the derivatives of $\ln D_N[f](\lambda)$, and then doing the resulting contour integrals, Keating and Mezzadri showed the following:\\
\emph{$\UN$ group/TI chains:} 
\begin{align}
S(N) = \frac{R}{3} \log N + \frac{R}{3}(K/\ln2 -6I_3) +o(1). \label{UNent}
\end{align}
\emph{Other groups/chains with boundary:} 
\begin{align}
S(N) =  \frac{R}{6} \log 2N + \frac{R}{6}(K/\ln2 -6I_3) +o(1).\label{ONent}
\end{align}
In these formulae: \begin{equation} \label{K}
K = 1 + \gamma_{E} + \frac{1}{R}\left(\sum_{r=1}^R\ln\abs{1
- \rme^{\rmi 2 \theta_r}} -2\sum_{1 \le r < s \le R} (-1)^{(r +
s)} \ln \left |\frac{1 - \rme^{\rmi\left(\theta_r -
\theta_s\right)}}{1 - \rme^{\rmi \left(\theta_r +
\theta_s\right)}}\right| \right),
\end{equation}
$\gamma_E\simeq0.58$ is the Euler-Mascheroni constant, and $I_3 \simeq 0.022$ is a definite integral evaluated in \cite{JK}.\par
We can see that \eqref{UNent} and \eqref{ONent} exactly verify the CFT predictions \eqref{UNpre} and \eqref{ONpre}. In particular we note that the boundary entropy is confirmed to be zero for all of these models, and we see in $\eqref{K}$ the dependence of the constant on fine details of the dispersion relation -- i.e. the values of $\{\theta_i\}$.
\subsection{Correlation functions}
We will now compute the ground state correlators using the Fisher-Hartwig conjecture. For the $\sigma_z$-string correlator we review the correct calculation given in \cite{HKM} for the TI chain, as well as giving a partial result for the extension to systems with a boundary. For the $\sigma_x$ and $\sigma_y$ correlators we amend an important detail of the TI calculation in \cite{HKM} which corrects the constant found there, and discuss why a similar extension to the chains with a boundary is not straightforward. We note that the correlators for the XX model were evaluated using the Fisher-Hartwig conjecture in \cite{Ov} -- our results agree when specialised to this case.\par
The general method in this section is given in \cite{LSM}. To summarise: we can rewrite, say, a two point spin-spin correlation function by the Jordan-Wigner transform as a fermionic $O(N)$-point function. As we have a quadratic theory in the fermions, we can use Wick's theorem to expand this as a determinant of a matrix of fermionic correlators. We know these fermionic correlators and so may then use the Fisher-Hartwig conjecture and related techniques to find the asymptotics of the spin-spin correlation functions.\par

\subsubsection{$\left\langle \prod^{N-1}_{i=0} \sigma^{z}_{i}  \right\rangle$}
The expectation value of the $\sigma^z$-string operator in the ground state is given by:
\begin{equation}
\left\langle \prod^{N-1}_{i=0} \sigma^{z}_{i}  \right\rangle = \det \mathcal{M}_{N} \label{ZCOR}
\end{equation}
where the matrix $\mathcal{M}_{N}$ is defined in Table \ref{models} for each of the different models. We use this for illustration as it is the simplest Fisher-Hartwig type correlator to calculate. The field theory prediction will be as in \eqref{vertexspinop}, except as $\sigma_z$ does not transform under the $\mathrm{U}(1)$ symmetry we must have that the operator corresponds to no overall charge creation -- i.e. $\sum\tilde\beta_j = 0$. This does not affect the leading order scaling which is $N^{-R/2}$.\\
\emph{U(N) group/TI chain:}\par For these systems, the matrix $\mathcal{M}_{N}$ is Toeplitz and thus we can directly apply Theorem \ref{FHT}. We find there are many representations for the symbol \eqref{symb} of this matrix, but since it is of the form described in Section \ref{toe}, those that are minimal will have $\tilde\beta_{j} = \pm \frac{1}{2}$ and $n_{j} \in\{1,-1,0\}$ along with constraint \eqref{nj}.
In order for $V(z)$ to satisfy the smoothness requirements we additionally require that
$\sum^{2R}_{r=1}  \beta_{r} = 0,$
then the symbol can be written in the form given by \eqref{fz} with
\begin{equation}
V(z) = V_0= \rmi \left(\sum^{R}_{r=1} \theta_{r} \left( \beta_{r} - \beta_{2R+1-r} \right) + 2 \pi \sum^{2R}_{r=R+1}  \beta_{r}\right)+(\rmi\pi)
\end{equation}
where we have used the evenness of the symbol. The final term $\rmi\pi$ fixes the correct overall sign of the symbol at zero, so is included here if and only if the symbol is negative at zero. \par
There are ${2R}\choose{R}$ different combinations $\tilde\beta_j = \pm 1/2$ which satisfy the constraints, thus there are ${2R}\choose{R}$ different representations with minimal weight. The labels $\beta,\tilde\beta$ and $n$ are fixed by an arbitrary initial representation $\{\beta_j\}$, but the final answer is independent of this choice.\par
Theorem \ref{FHT} then gives us the following leading asymptotic behaviour in the limit $N \rightarrow \infty$:
\begin{align}\label{corrz}
\nonumber \left\langle \prod^{N-1}_{i=0} \sigma^{z}_{i}  \right\rangle &  \sim N^{- \frac{R}{2}} (-1)^{RN}  \left( G \left( \frac{3}{2} \right) G \left( \frac{1}{2} \right) \right)^{2R} \\
&  \sum_{\mathrm{Reps:}~\{\tilde\beta_j\}} \exp{\rmi N \sum^{R}_{r=1} \theta_{r} \left(  \tilde{\beta}_{r} - \tilde{\beta}_{2R+1-r}  \right)} \prod_{1 \leq j<k \leq 2R} |\exp(\rmi\theta_j)-\exp(\rmi\theta_k) |^{2 \tilde{\beta}_j \tilde{\beta}_k}.
\end{align} Here we did not include the extra $\rmi \pi$, i.e. we assume the symbol is positive at zero -- this term would give an additional coefficient $(-1)^N$ if included. This is in exact agreement with the field theory calculation, and gives us the coefficients. We also note that by translation invariance we can take any site as the starting point of our string.\\
\emph{Other groups/chain with boundary:} \par 
For systems with a boundary, corresponding to the other classical compact groups, we still have \eqref{ZCOR} -- but now the matrix ${\mathcal{M}}_N$ has a Toeplitz+Hankel structure as dictated by Table \ref{models}. The method to find asymptotics of such matrices is outlined in \cite{DIK} -- they are transformed first to Hankel matrices and then to Toeplitz matrices multiplied by certain orthogonal polynomials. Using the results of the paper, the asymptotics of the Toeplitz part is controlled by Theorem \ref{FHT}, and the asymptotics of the orthogonal polynomials are derived by the Riemann-Hilbert method, a discussion of which is beyond the scope of this paper. \par
A general theorem for Toeplitz+Hankel matrices with symbols containing jump discontinuities is not given in \cite{DIK} -- however combining \cite[Theorem 1.25 and Remark 1.22]{DIK}, we have the following:
\begin{align}
 \left\langle \prod^{N-1}_{i=0} \sigma^{z}_{i}  \right\rangle &   \sim N^{- \sum \beta_j^2}(A\exp(\rmi F(N)) +o(1))=
N^{- \frac{R}{2}}(A\exp(\rmi F(N))+o(1)).
\end{align}
$A$ is an unknown constant, and $F$ is an unknown real function. Both could in principle, however, be determined rigorously according to the arguments of \cite{DIK}. In particular, the additional Riemann-Hilbert analysis required due to our symbol violating the hypotheses of \cite[Theorem 1.25]{DIK} only affects the coefficient, and not the leading order algebraic decay. We also note in the derivation of \cite[Theorem 1.25]{DIK}, and in the theorem itself, the determinant scales as $(2N)^{-R/2}$ -- this is another clear connection with the method of images and conformal field theoretic calculations where we would expect to make such a replacement.
This reasoning applies to all the models with boundary that we consider, and in each case we get the bulk scaling as we would expect from the field theory.\par
We note that for systems with these Toeplitz+Hankel structures we lack translation invariance, so this result is restricted to the $\sigma^{z}$-string on the first $N$ sites.

\subsection{$\left\langle \sigma^{y}_{0} \sigma^{y}_{N} \right\rangle$}

The ground state $\sigma_y$ two-point function has the form
\begin{equation}
\left\langle \sigma^{y}_{0}  \sigma^{y}_{N}  \right\rangle = \det \left( \mathcal{M}_{j+1,k} \right)_{j,k=0}^{ N-1} = \det \left( \mathcal{M}^{y}_{j,k} \right)_{j,k=0}^{N-1},
\label{yy} 
\end{equation}
where the matrix $\mathbf{\mathcal{M}}^{y}$ has symbol 
\begin{equation}
f^{y} \left( \theta \right) = e^{ -\rmi \theta} \frac{\Lambda (\theta)}{|\Lambda(\theta)|}.
\end{equation}
\emph{U(N) group/TI chain}\par
Again for the translation invariant system, the matrix $\mathbf{\mathcal{M}}^{y}$ is Toeplitz and we can use Theorem \ref{FHT}. This results in the same analysis as for the previous correlator and we obtain asymptotics of the same form as the right hand side of \eqref{corrz}. The only difference is that the condition on the set of $\{\beta_j\}$ is
\begin{equation}
\sum^{2R}_{r=1}\beta_{r} =-1,
\label{by}
\end{equation}
and we find the $\{\tilde\beta_j\}$ subject to the same constraint \eqref{nj}. One must also check whether to include the additional $\rmi\pi$ in $V_0$ -- here and below it corresponds to a positive symbol at zero. This gives ${2R}\choose{R+1}$ different Fisher Hartwig representations to sum over. The condition \eqref{by} on the $\{\beta_j\}$ is all we need to correct the calculation in \cite{HKM}. By translation invariance this extends to $\left\langle \sigma^{y}_{i}  \sigma^{y}_{i+N}  \right\rangle$. \\
\emph{Other groups/chain with boundary}\par
From Table \ref{models} we see that we have a particular Toeplitz+Hankel matrix to deal with. However, to use the methods of \cite{DIK} we need to shift the indices on the matrices as above by multiplying the symbol by an exponential. This means that the symbol is no longer even in $\theta$ so we can no longer use the results of \cite{DIK} for Toeplitz+Hankel matrices. The same issue affects the $\sigma_x$ correlator. 

\subsection{$\left\langle \sigma^{x}_{0} \sigma^{x}_{N} \right\rangle$}\par
Finally we discuss the ground state $\sigma_x$ two-point function:
\begin{equation}
\left\langle \sigma^{x}_{0}  \sigma^{x}_{N}  \right\rangle = \det \left( \mathcal{M}_{j,k+1} \right)_{j,k=0}^{N-1}.
\label{xxcorr}
\end{equation}
\emph{U(N) group/TI chain}\par
We may use Theorem \ref{FHT} by rewriting $\left( \mathcal{M}_{j,k+1} \right)_{j,k=0}^{N-1}$ as $ \left(\mathcal{M}^{x}_{j,k} \right)_{j,k=0}^{N-1}$ where matrix ${\mathcal{M}}^{x}$ has symbol:
\begin{equation}
f^{x} \left( \theta \right) = e^{ \rmi \theta} \frac{\Lambda (\theta)}{|\Lambda(\theta)|}.
\label{gx}
\end{equation}
The analysis for the Toeplitz determinant with symbol \eqref{gx} is again the same as for the previous cases, with the asymptotics taking the form of the right hand side of \eqref{corrz}, but with the condition: 
\begin{equation}
\sum^{2R}_{r=1} \beta_{r} =1. \label{bx}
\end{equation}
Again we have ${2R}\choose{R+1}$ different possible representations and by translation invariance we have $\left\langle \sigma^{x}_{i}  \sigma^{x}_{i+N}  \right\rangle$.\par
We remark that the form of the right hand side of \eqref{corrz} combined with the conditions \eqref{by} and \eqref{bx} gives us that the $\sigma_x$ and $\sigma_y$ correlators are equal -- as we expected from rotation invariance. We see agreement with the field theory calculation for both of these correlators.
We now illustrate how to use the above results in a simple case:

\subsection{Example: $\left\langle \sigma^{x}_{i}  \sigma^{x}_{i+N}  \right\rangle_{}$ and $\left\langle \sigma^{y}_{i}  \sigma^{y}_{i+N}  \right\rangle_{}$ for  $R=2$}
We consider any dispersion which has two zeroes of $\Lambda(\theta)$ in $(0,\pi)$, with $\Lambda(0)>0$. This could be, for example, a model with positive nearest neighbour hopping on the even and odd sublattices. We find $V(z)$ takes the form
\begin{equation}
V(z) = \rmi \left(\theta_{1} \left( \beta_{1} - \beta_{4} \right) + \theta_{2} \left( \beta_{2} - \beta_{3} \right) \right) + 2 \pi\rmi \left( \beta_{3} + \beta_{4} \right) + \rmi\pi, \label{veq}
\end{equation}
and we have the following four representations which satisfy the constraints: \\
\bgroup
\def\arraystretch{1.5}
\begin{figure}[h]
\centering
\begin{tabular}{c c c c c c c}
\hline \hline
 & $f(z;n_{1},n_{2},n_{3},n_{4})$ & $\tilde\beta_{1} $ & $\tilde\beta_{2}$ & $\tilde\beta_{3}$ & $\tilde\beta_{4}$  \\
 \hline\hline
 Rep 1 & $f(z,0,0,0,0)$ & ${1/}{2}$ & $1/2$ & $1/2$ & $-1/2$  \\
 Rep 2 & $f(z;0,0,-1,+1)$  & $1/2$ & $1/2$& $-1/2$& $1/2$  \\ 
 Rep 3 & $f(z,-1,0,0,+1)$ & $-1/2$ & $1/2$ & $1/2$ & $1/2$\\
 Rep 4 & $f(z;0,-1,0,+1)$ & $1/2$ & $-1/2$ & $1/2$ & $1/2$& \\
  \hline \hline 
\end{tabular}.
\end{figure}\egroup\\
Summing over the contribution of these representations as in \eqref{corrz} gives the following result:
\begin{align}
\nonumber\left\langle \sigma^{x}_{i} \sigma^{x}_{i+N} \right\rangle \sim&~  (-1)^NN^{-1} \left( G\left(\frac{3}{2}\right) G\left(\frac{1}{2}\right) \right)^{4} \\&\times\left( \sqrt{\frac{\sin \theta_{2}}{  \sin \theta_{1}} } 2\cos \left( N \theta_{1} \right) + \sqrt{\frac{\sin \theta_{1}}{  \sin \theta_{2}} } 2\cos \left( N \theta_{2} \right) \right). \label{xex}
\end{align}
The $\sigma_y$ calculation is very similar. We have the same formula for $V$, but the signs of $n$ and $\tilde\beta$ the representations are reversed in the table above. Doing the sum we get identical asymptotics to \eqref{xex} as expected.
\section{Conclusions and discussion}
In this paper we reconciled a rigorous mathematical approach to calculating properties of spin chains with the intuition gained from more heuristic physical arguments. To do this we worked within a framework of mathematically tractable models introduced in \cite{KM}. We motivated an effective CFT description of each of these models which gave predictions consistent with the rigorous calculations. This was particularly informative after noting that some of the models in \cite{KM} represent systems with a boundary. We then gave the results of rigorous calculations in \cite{KM} for the entanglement entropy, and \cite{HKM} for the translation invariant correlation functions. We fixed an error in \cite{HKM} and gave a partial extension to the expectation of the $\sigma_z$ string operator for systems with a boundary. All of these agreed with the field theory expectation. We could not directly extend this method in the presence of a boundary to the more physically relevant spin-spin correlation functions, but we note the formal similarity between the Jordan-Wigner transformed $\sigma_x$ and $\sigma_y$ operators to the $\sigma_z$ string. It is interesting to compare the partial result \eqref{fieldcorr} with the full lattice results. Given a more thorough field theoretic calculation perhaps the entire Fisher-Hartwig result could be recovered -- for example we might expect the combinations of sine functions in \eqref{xex} to come from the theory of form factors in integrable quantum field theories \cite{KBI,Smi}. The universal appearance of the combination $\left(G(1/2)G(3/2)\right)^2$ for each complex fermionic field suggests this is simply a normalisation of the field operators relative to the normalisation chosen on the lattice. We also remark that the oscillating prefactor $(-1)^N$ which we included in \eqref{vertexspinop} in an ad-hoc manner is found to be present\footnote{After fixing a lattice Jordan-Wigner convention that corresponds to anti-ferromagnetic order.} if and only if there is a difference ($\pm1$) in the number of fermion types with positive $v_i$ and those with negative $v_i$ in $(0,\pi)$. If we imagine a `Dirac cone' where we extend the linearised dispersion to an intersection at $\theta=0$ in each fermionic sector, the sign of $v_i$ corresponds to a vacuum filled with either particles or holes. This picture could be a way of fixing the correct continuum spin operator, although we could not see how to apply it directly in our heuristics.
\par
To summarise the universality results, we see that for our critical systems in both the effective theory and the Fisher-Hartwig calculation, the most important feature affecting the physics is the number of zeroes of the dispersion relation. Interestingly, both calculations are sensitive to the values of the zeroes, or Fermi momenta, of the dispersion, but the gradient at these zeroes, or the Fermi velocities, are invisible to the Fisher-Hartwig analysis. They do, however, feature prominently in the physics -- in particular the spacetime mixing conformal symmetries must act on coordinates $(x,v_{\mathrm{F}} t)$. For this reason, attempting to extend this analysis to dynamic correlators would be an interesting question. We note that from the boundary CFT perspective, for correlation functions between sites at the boundary (but displaced in time) we would expect different scaling exponents to the bulk correlators. There is also an interesting question about the enhanced symmetry for systems where the Fermi velocity is the same at two or more crossings. This system has a new symmetry at the effective theory level -- we can now mix these different fermion sectors -- however the lattice results imply that the physics does not change in this case.\par
To address universality more precisely, we have rigorously demonstrated different classifications depending on the macroscopic quantity of interest as follows: 
\begin{itemize}
\item {Classification }I:  The divergence of the entanglement entropy of a block for translation invariant periodic systems of free fermions which preserve fermion number (and related spin chains) is universal for systems with the same number of zeroes of their dispersion.  
\item {Classification Ia:} The divergence of the entanglement entropy at the boundary for free fermion systems with a boundary which preserve fermion number and which have a translation invariant bulk (and related spin chains) is universal for systems with the same number of zeroes of their dispersion.  
\item {Classification II, IIa}: The subclass of such systems sharing entanglement entropy up to $O(1)$ is given by systems with equal Fermi momenta $\{\theta_i\}$. 
\item The leading asymptotics of the equal time spin-spin correlation functions for the spin chains in Class I have universal scaling and are universal for Class II.
\end{itemize} \par
We now make some general remarks on the mathematical methods and further applicability. The Fisher-Hartwig conjecture as proved in \cite{DIK} relies on a matrix with Toeplitz structure or some special Toeplitz+Hankel structures. In the approach above, these structures correspond to bulk translation invariance. If we restrict to free fermion models (and related spin chains), the class of microscopic systems we have studied effectively saturates those with bulk translation invariance. In particular, the finite chain must be either a ring or have two open ends and our analysis covers both in the thermodynamic limit. We would thus need some other rigorous arguments to extend the microscopic universality classes to systems without translation invariance, but which we would expect to have the same universal behaviour based on an effective description. 
Another open problem would be moving the analysis away from the edge in the case of the open chain -- this would require Toeplitz+Hankel results beyond those given in \cite{DIK}.
The rigorous analysis above also relied heavily on the fact that our system had a description as a quadratic fermionic Hamiltonian. This allowed us to write the entanglement entropy in terms of eigenvalues of the correlation matrix, and also allowed us to use Wick's theorem to compute correlation functions. Hence we cannot expect to use Fisher-Hartwig type asymptotics directly in interacting systems.\par
We note the link between matrix Riemann-Hilbert problems, via the results of \cite{DIK}, and rigorously providing asymptotics for systems with a CFT description. This may be unsurprising when we consider the strong links to the theory of differential equations from both sides. Other interesting results would follow from solving related Riemann-Hilbert problems. For example it should also be possible to use the ideas contained in \cite{CW} to understand finite size effects for our lattice model -- i.e. studying long distance correlation functions before taking the limit $M\rightarrow\infty$. In \cite{CK} and \cite{CIK} uniform asymptotics are obtained for two merging Fisher-Hartwig singularities and a single emerging singularity respectively, giving information about the phase transition between different asymptotic behaviours. The phase transitions in our model, due to the form of the symbol, correspond to two singularities emerging at the same time -- hence these results are not directly applicable. It would thus be another interesting open question to study the case of the simultaneous emergence of two Fisher-Hartwig singularities via the Riemann-Hilbert method.  \par
Finally we remark that we have far from utilised all of the symbols amenable to Fisher-Hartwig analysis. It would be interesting to see if any other spin chain models could be connected to symbols with different Fisher-Hartwig singularity structures, or indeed if in general there is a corresponding physical model for each symbol. We would expect a connection to Bethe ansatz solvable models through the methods contained in \cite{KBI}. 
\subsection*{Acknowledgements}
We gratefully acknowledge E. Casali, F. Flicker, J. Griffin, J. H. Hannay, F. Mezzadri and J. M. Robbins for helpful discussions. We thank J. P. Keating and R. Verresen for crucial remarks as well as comments on the manuscript. JH gratefully acknowledges support from the Leverhulme Trust during the final stages of this work.\bibliography{spins}{}
\bibliographystyle{JHEP}
\appendix
\section{Treatment of points where linearisation fails}\label{exc}
We need to consider what happens at an exceptional point $\theta_i$ where $\Lambda(\theta_i)=0$ and  $\Lambda'(\theta_i)=0$.
At such points we \emph{may} get different critical behaviour. For example in the XX model in a magnetic field, $|\Lambda(q)|=|\cos(q)+h|$. The model is gapped for $|h|>1$, has two crossings and is linearisable when $|h|<1$, but when $h=1$ it has a low energy parabolic dispersion (about $q=\pi$) and the effective theory is that of a non-relativistic massless fermion, which is not a CFT. \par
We note that the rigorous argument based on the Fisher-Hartwig conjecture is only sensitive to the following question: does the sign of the dispersion change at the crossing? If it does - i.e. if the first non-zero derivative is odd --  then we get identical Fisher-Hartwig behaviour to a linear crossing at the same point. This can be understood physically in the following way: an infinitesimal perturbation $\Lambda \rightarrow \Lambda \pm\epsilon$ will now have a linear crossing in the neighbourhood of $\theta_i$. There is thus stability here -- infinitesimally either side of our exceptional parameters we have a system with $n$ linear crossings -- hence we should expect by continuity that such an effective description still applies at such exceptional points. \par
In the case that the first non-zero derivative is even, we have an instability of the type described above for the XX model. An infinitesimal shift changes the number of crossings, and hence this exceptional point is at a phase transition between different critical theories. Due to the nature of this singularity, the symbol cannot be written in the general Fisher-Hartwig form \cite{DIK} at this point, so that form of analysis fails. Hence we do not seek to describe these transition points in the same terms, and exclude such exceptional $\Lambda(q)$ from our discussion. We note that this automatically excludes any zeroes at $q=0$ and $q=\pi$ by symmetry.

\section{Method of images for the chain with boundary and relation to classical compact groups}\label{images}
\subsection{$\SpN$}
Consider the model on the half chain:
\begin{align} 
H = \sum_{j,k=1}^{\infty} a(j-k) B^\dagger_jB_k - \sum_{j,k=1}^{\infty} a(j+k)B^\dagger_jB_k,\label{neg}
\end{align}
where $a(j) = a(-j)$ and decays for large $j$. 
We will show that:
\begin{itemize}
\item this model is solvable by the method of images,
\item this can be written as an $\SpN$ model in the language of \cite{KM}.
\end{itemize}
To do the first, let us define $B_{-j} := - B_j$. Then, as in \ref{ord}, we can write:
\begin{align}
H &= \frac{1}{2}\left(\sum_{j,k=1}^{\infty} a(j-k) B^\dagger_jB_k+\sum_{j,k=1}^{\infty} a(j-k) B^\dagger_{-j}B_{-k}\right) - \sum_{j,k=1}^{\infty} a(j+k)B^\dagger_jB_k\nonumber \\
&= \frac{1}{2}\left(\sum_{j,k=1}^{\infty} a(j-k) B^\dagger_jB_k+\sum_{j,k=1}^{\infty} a(j-k) B^\dagger_{-j}B_{-k}\right) + \sum_{j,k=1}^{\infty} a(j+k)B^\dagger_jB_{-k} \nonumber\\
&= \frac{1}{2}\sum_{j,k=-\infty}^{\infty} a(j-k) B^\dagger_jB_k. \label{ordti} \end{align}
This is now a TI model which can be solved by Fourier transform.
\par
Now to relate this to $\SpN$, we define $B_j = \frac{b_j-b_{-j}}{\sqrt 2}$. Then:
\begin{align}
H& = \frac{1}{2}\left(\sum_{j,k=1}^{\infty} a(j-k) (b_j^\dagger-b_{-j}^\dagger)(b_k-b_{-k}) - \sum_{j,k=1}^{\infty} a(j+k)(b_j^\dagger-b_{-j}^\dagger)(b_k-b_{-k}) \right)\nonumber\\
& = \frac{1}{2}\left(\sum_{j,k=1}^{\infty} (a(j-k) - a(j+k)) (b_j^\dagger b_k+b_{-j}^\dagger b_{-k}-b_j^\dagger b_{-k}-b_{-j}^\dagger b_{k}) \right)\nonumber\\
& =\frac{1}{2} \sum_{j,k=-\infty}^{\infty} (a(j-k) - a(j+k)) b_j^\dagger b_k \label{sp}
\end{align}
As in section \ref{ord}, we need to shift the lattice by one site to see that this is our model \eqref{ham} with $\overline A_{jk} = a(j-k) - a(j+k+2)$. Hence we see that the half-line model \eqref{neg} is the $\SpN$ model of \cite{KM} in full generality. Note that \eqref{sp} may be transformed to \eqref{ordti} due to the equivalence $B_j= -B_{-j}$ which carries over to the Fourier transformed operators $B_{q}=-B_{-q}$.  This means any eigenstate $B_q^\dagger \ket{0}$ will have an eigenvalue of $2\Lambda(q)$ where $\Lambda(q)$ is the dispersion one would get by naively applying the diagonalisation in \eqref{TI} to \eqref{ordti}. Equivalently one can write the Hamiltonian purely in terms of the independent $k$ modes, evenly spaced in $(-\pi,\pi]$, and send $\Lambda(q) \rightarrow 2\Lambda(q)$ in the process. This doubled dispersion is seen directly in the diagonalisation carried out in \cite{KM} where the same modes are written twice as densely in the interval $[0,\pi)$. We prefer $(-\pi,\pi]$ as it gives the same density of states as the translation invariant model when counting the gapless degrees of freedom.  We remark that the initial model \eqref{neg} is the natural physical interpretation of all three models in this section, as it is the only one which is local in the thermodynamic limit. 
\subsection{$\ONen$ and $\ONo{+}$}
The structure of the $\overline A$ matrix for the group $\ONen$ is identical to that for $\SpN$ so all of the above applies in exactly the same way. The fermion chain related to $\ONo{+}$ 
is structurally very similar, but naturally lives on the half integer sites, i.e. we should shift the labels by $j \rightarrow j' = j+1/2$ and then repeat the above calculation for \begin{align} 
H = \sum_{j,k=0}^{\infty} a(j-k) B^\dagger_{j+1/2}B_{k+1/2} - \sum_{j,k=0}^{\infty} a(j+k)B^\dagger_{j+1/2}B_{k+1/2}.
\end{align}
Alternatively, the $\ONo{+}$ chain can be embedded in the $\SpN$ chain by shifting the chain by half a lattice space  and then relabelling sites by integers $j'\rightarrow 2j'$. Both are consistent with the diagonalisation given in \cite{KM}. As we can embed the chain corresponding to each of these groups in the $\SpN$ chain, we do not extend the class of models by including them.
\subsection{$\ONep$ and $\ONo{-}$}
One can repeat the above manipulations for the following Hamiltonian on the half chain (starting at zero):
\begin{align} \label{pos}
H = \sum_{j,k=0}^{\infty} a(j-k) B^\dagger_jB_k + \sum_{j,k=1}^{\infty} a(j+k)B^\dagger_jB_k.
\end{align}
We put $B_{-j}:=B_j$ to reach a translation invariant model, and $B_j = \frac{1}{\sqrt{2}}(b_j+b_{-j})$ to reach the $\ONep$ model of \cite{KM}. The only novelty is that the site at zero now has $B_0 \neq 0$, and we must include $B_{-0}=B_0$. This is seen in the $\ONep$ model where the operator $b_0= \sqrt 2 B_0$ has a heavier weight in calculations (see Table \ref{models}). We also note that this subtlety is seen at the level of the dispersion as well: we should not send the TI $\Lambda(0)\rightarrow 2\Lambda(0)$ for the $\{B_k\}$ model \eqref{pos}, but it does appeared doubled in the $\{b_k\}$ $\ONep$ model.\par
We may also send $j \rightarrow j +1/2$ and consider the model:
\begin{align} \label{pos}
H = \sum_{j,k=0}^{\infty} a(j-k) B^\dagger_{j+1/2}B_{k+1/2} + \sum_{j,k=0}^{\infty} a(j+k)B^\dagger_{j+1/2}B_{k+1/2}.
\end{align} Again we obtain a TI model after reflection and, with $B_j = \frac{1}{\sqrt{2}}(b_j+b_{-j})$, we reach the  $\ONo{-}$ model. As there is no site reflected into itself, we do not get any special weights. These are again in agreement with the diagonalisation in \cite{KM}.\par
We note that the connected subgroups of the orthogonal groups containing the negative of the identity matrix,$-\mathbb{I}$, correspond to systems at the extraordinary transition, and those not containing $-\mathbb{I}$ to systems at the ordinary transition. Finally we remark that by the addition of a second auxiliary site between the reflected site $-M$ and the boundary site $M$, the method of images can be used for a finite system with two boundaries. Care should be taken with the quantised momentum for the finite system, but the manipulations are otherwise identical. This method is used in \cite{FC} to study the XX model on a finite chain.

\end{document}